\documentclass{nature}
\usepackage{graphicx}
\makeatletter
\let\saved@includegraphics\includegraphics
\AtBeginDocument{\let\includegraphics\saved@includegraphics}
\renewenvironment*{figure}{\@float{figure}}{\end@float}
\makeatother
\usepackage{xcolor}
\usepackage[version=4]{mhchem}
\usepackage{scicite}
\usepackage{mathtools}
\usepackage{times}
\usepackage{amsmath,amssymb,colordvi,mathrsfs,bm,verbatim,dcolumn,epsfig,subfigure,float}

\usepackage{hyperref} 
\usepackage{cleveref} 

\hypersetup{colorlinks = true, 
	    linkcolor = blue, 
	    urlcolor = blue,
            citecolor = blue} 

\title{Orthogonal Geometry of Magneto-Optical Kerr Effect Enabled by Magnetization Multipole of Berry Curvature}

\newcommand{\ICQD}{1}
\newcommand{\UESTC}{2}
\newcommand{\MicroElec}{3}
\newcommand{\Phys}{4}
\newcommand{\NSRL}{5}
\newcommand{\HighField}{6}
\newcommand{\FudanFICS}{7}

\author{Haolin Pan$^{\ICQD}$, Han Li$^{\UESTC}$, Jixiang Huang$^{\MicroElec}$, Zheng Liu$^{\Phys}$, Mingyue Fang$^{\ICQD}$, Yanan Yuan$^{\NSRL}$, Daxiang Liu$^{\NSRL}$, Xintong Hu$^{\HighField}$, Wenzhi Peng$^{\ICQD}$, Zhenguo Liang$^{\ICQD}$, Xiao Chang$^{\HighField}$, Zhigao Sheng$^{\HighField}$, Xianzhe Chen$^{\FudanFICS}$, Lingfei Wang$^{\ICQD}$, Qian Li$^{\NSRL}$, Peng Li$^{\MicroElec}$, Qian Niu$^{\Phys}$, Yang Gao$^{\ICQD,\Phys}$, Qinghui Yang$^{\UESTC}$, Dazhi Hou$^{\ICQD,\Phys\ast}$\\
\\
\normalsize{$^{\ICQD}$International Center for Quantum Design of Functional Materials, Hefei National Research Center for Physical Sciences at the Microscale, University of Science and Technology of China, Hefei 230026, China.}\\
\normalsize{$^{\UESTC}$School of Electronic Science and Engineering, University of Electronic Science and Technology of China, Chengdu 611731, China.}\\
\normalsize{$^{\MicroElec}$School of Microelectronics, University of Science and Technology of China, Hefei 230026, China.}\\
\normalsize{$^{\Phys}$Department of Physics, University of Science and Technology of China, Hefei 230026, China.}\\
\normalsize{$^{\NSRL}$National Synchrotron Radiation Laboratory, University of Science and Technology of China, Hefei 230026, China.}\\
\normalsize{$^{\HighField}$High Magnetic Field Laboratory, Chinese Academy of Sciences, Hefei, 230031, China.}\\
\normalsize{$^{\FudanFICS}$Frontier Institute of Chip and System, Fudan University, Shanghai 200000, China.}\\
\normalsize{$^\ast$Corresponding author. Email: dazhi@ustc.edu.cn(D.H.)}
}

\begin{document} 


\baselineskip24pt


\maketitle 
\newpage
\begin{abstract}
\
\\
The Magneto-Optical Kerr Effect (MOKE) is a fundamental tool in magnetometry, pivotal for advancing research in optics, magnetism, and spintronics as a direct probe of magnetization. 
Traditional MOKE measurements primarily detect the magnetization components parallel to the Poynting vector, which can only access the magnitude but not the direction of the orthogonal component.
In this study, we introduce an orthogonal MOKE geometry in which the Kerr signal detects both the magnitude and direction of the magnetization component perpendicular to the Poynting vector.
We demonstrate the broad applicability of this orthogonal geometry through the MOKE measurements in cubic ferromagnets and van der Waals ferromagnet. We theoretically show that the orthogonal MOKE geometry is enabled by the multipolar structure of Berry curvature in the magnetization space, which generally induces a Voigt vector orthogonal to the magnetization, thereby accounting for the unique magnetization angle dependence distinct from conventional MOKE. The establishment of the orthogonal MOKE geometry not only introduces a new paradigm for magneto-optical measurements but also provides a framework for exploring the magnetization multipoles of Berry curvature across the electromagnetic spectrum.
\end{abstract}
\newpage
The Magneto-optical Kerr Effect (MOKE) is a vital technique for magnetization sensing, with extensive applications in both research and industry \cite{2000_Qiu_MOKEreview, 1984_MODisc, 2003_MOrecording}. 
Traditionally, magnetization-odd Kerr rotation or ellipticity of light can be observed from ferromagnets via two MOKE geometries: polar MOKE and longitudinal MOKE, as illustrated in Fig. \ref{fig1}a and \ref{fig1}b \cite{1877_Kerr, 1878_Kerr, 1954_Edward_LMOKE, 2000_Qiu_MOKEreview}. 
The Voigt vector $\boldsymbol Q$, which describes the Kerr rotation and ellipticity, is commonly assumed parallel to the magnetization $\boldsymbol M$\cite{1968_Freiser_MOreview, 1990_Zak_MO, 1997_Zvezdin_Book}. 
The $\boldsymbol Q \parallel \boldsymbol M$ alignment between the Voigt vector and the magnetization dictates that the MOKE can only detect the magnetization component parallel to the Poynting vector as in the longitudinal and polar geometry of MOKE\cite{2000_Qiu_MOKEreview}. 
This situation mirrors the collinear relationship between magnetization and Berry curvature in the anomalous Hall effect, positioning the MOKE as the optical analogue of the Hall effect\cite{2006_Nagaosa_AHE, 2010_MacDonald_AHE, 2015_Hou_AHE}. 
Consequently, the magnetization orthogonal to the light has generally been considered beyond the access of the MOKE. 
Although rare exceptions have been observed in crystalline samples with vicinal surfaces \cite{1999_Rasing_Vicinal, 2003_Hamrle_Vicinal}, the required low symmetry can be rarely satisfied in most ferromagnets, precluding the vicinal interface sensitive MOKE from being accepted as a general MOKE geometry.

The recent observation of the in-plane anomalous Hall effect in ferromagnets suggests that an out-of-plane Berry curvature can be induced by an in-plane magnetization\cite{2024_WU_IPAHE,2024_PWZ_IPAHE}. 
This unconventional geometry can be comprehended through the multipolar structure of Berry curvature in magnetization space, which was proposed as an analogy to Berry curvature multipoles in momentum space \cite{2024_ZhengLiu_Multipolar,2019_nonlinearHall,2015_Fu_PRL,2021_Qiong_NM}. Critically, this out-of-plane Berry curvature under in-plane magnetization also induces an out-of-plane $\boldsymbol Q$ component \cite{2004_Yao,2024_Xu_viewpoint}, which challenges the $\boldsymbol Q \parallel \boldsymbol M$ assumption in traditional MOKE analysis and should enable the detection of the magnetization component orthogonal to the Poynting vector. Specifically, the out-of-plane $\boldsymbol Q$ can generate a magnetization-odd MOKE signal in light reflected normally off a ferromagnet with in-plane magnetization as illustrated in Fig. \ref{fig1}c, thereby realizing the orthogonal MOKE geometry. This mechanism is also valid in ferromagnetic insulators, where $\boldsymbol Q$ can remain finite at optical frequencies even in the absence of conduction electrons. Crucially, if the Voigt vector indeed reflects the multipolar structure of Berry curvature in magnetization space, which is argued by our theory to be ubiquitously present in ferromagnets \cite{2024_ZhengLiu_Multipolar}, the orthogonal MOKE geometry should not be limited to specific systems. Instead, it may represent a general MOKE geometry applicable across a wide range of ferromagnets, a prospect yet to be fully explored.

In this work, we demonstrate the orthogonal MOKE geometry both in conventional cubic ferromagnets and van der Waals ferromagnet, representing the majority of ferromagnetic materials studied in magneto-optics, thereby verifying the broad applicability. Before showing the experimental results, we first use the ferromagnet with cubic lattice as an example to derive the relation between the magnetization multipole of Berry curvature and a Voigt vector orthogonal to magnetization. Following the theoretical work \cite{2024_ZhengLiu_Multipolar}, we apply the multipolar expansion to $\boldsymbol Q$ in magnetization $\boldsymbol M$ space, which can be simplified in the cubic lattice as follows: 
\begin{equation}
	Q_{i} =\alpha m_i+\beta m^3_i+\cdots,
	\label{equ:cubic}
\end{equation}
considering the symmetry operations of cubic lattice. In this formula, $\alpha$ mainly comes from the dipole and the first term reproduces the familiar linear relation between $\bm Q$ and $\bm M$, while $\beta$ is dominated by the octupole and induces the leading-order nonlinearity between $\boldsymbol Q$ and $\boldsymbol M$. $\boldsymbol m$ denotes the unit vector of $\boldsymbol M$ in the cubic lattice coordinates.
A detailed discussion on the multipolar expansion and symmetry analysis is included in the section I of supplementary material.
To demonstrate the essential role of the multipole for the misalignment between $\boldsymbol M$ and $\boldsymbol Q$, which directly enables the orthogonal MOKE geometry, we compare the component of $\boldsymbol Q$ perpendicular to $\boldsymbol M$ contributed by the $\alpha m_i$ and $\beta m^3_i$ terms as examples, shown in Fig. \ref{fig1}e and \ref{fig1}f, respectively. 
The $\alpha m_i$ term is found to cause a perfect $\boldsymbol M$ $\parallel$ $\boldsymbol Q$ alignment so that \(\boldsymbol{Q_\bot}\) equals zero, while the $\beta m^3_i$ term introduces a finite \(\boldsymbol{Q_\bot}\) for most orientation of $\boldsymbol M$. For instance, the $\boldsymbol M$ pointing [112] can induce a $\boldsymbol Q$ along [118] via the $\beta m^3_i$ term. 
It is worth noting that such misalignment between $\boldsymbol Q$ and $\boldsymbol M$ can generally arise from the contributions of octupole and higher-order terms in the multipolar structure present in non-cubic lattice systems as well.

Our experimental investigations begin with the detection of the orthogonal MOKE in a 1-$\mu m$ $\rm Tb_{2.22}Bi_{0.78}Fe_{5}O_{12}$ (Tb:BIG) single crystal film with space group $Ia\bar{3}d$, which belongs to the garnet family widely used as magneto-optical materials \cite{2021_GarnetApplication , 2023_Faraday_isolator}.
Figure \ref{fig2}a depicts the experimental setup of the orthogonal MOKE measurement on the Tb:BIG(111) sample. 
A pronounced asymmetric signal with respect to \(\boldsymbol{H}\) is detected in both Kerr rotation $\theta_k$ and Kerr ellipticity $\varepsilon_k$ when $\varphi=0^\circ$, as shown in Fig. \ref{fig2}c, echoing the in-plane \(\boldsymbol{M}\)-\(\boldsymbol{H}\) curve of the sample in Fig. \ref{fig2}b. 
Since the Tb:BIG sample exhibits strong in-plane magnetic anisotropy with the saturation field below 50 Oe, the sudden jump in Kerr signal within 50 Oe is attributed to the in-plane rather than the out-of-plane magnetization. 
This result confirms that an \(\boldsymbol{M}\)-odd Kerr signal is generated in Tb:BIG by \(\boldsymbol{M}\) perpendicular to the light, which is in accordance with the orthogonal MOKE illustrated in Fig. \ref{fig1}c. 
The asymmetric components of the orthogonal MOKE signal, $\theta^A_k$ and $\varepsilon^A_k$, are quantified as $67.2\ \rm\mu rad$ and $52.4\ \rm\mu rad$, which are well above the measurement noise level of approximately $1\ \rm \mu rad$. It is worth noting that the Kerr rotation of the orthogonal MOKE almost amounts to 30\% of the longitudinal MOKE signal as shown in extended Fig. 2a, highlighting it as a substantial effect.

Figure \ref{fig2}d presents the \(\boldsymbol{H}\)-sweeping results of \(\theta_k\) when the in-plane orientation angle \(\varphi\) of the sample varies, showing that the \(\theta_k\) signal exhibits a three-fold symmetry.
Since the in-plane rotation of the sample does not alter the orthogonal geometry between \(\boldsymbol{M}\) and the probing light, we can conclude that the crystalline direction of \(\boldsymbol{M}\) in Tb:BIG determines the magnitude and sign of the orthogonal MOKE signal. 
The values of \(\theta^A_k\) at different \(\varphi\), shown in Fig. \ref{fig2}e, are accurately fitted with the function $a\cdot \sin(3\varphi)$, which is exactly the (111)-face contributions of octupole, 32-pole and 128-pole according to Eq. \ref{equ:cubic}, as detailed in Section II of the supplementary material. The \(\rm 120^\circ\)-periodic sinusoidal relationship is distinct from the familiar \(\rm 360^\circ\) period originating from the $\boldsymbol{M}$ dipole term in most MOKE measurements and rules out the vicinal surface as the origin\cite{1999_Rasing_Vicinal, 2003_Hamrle_Vicinal}. 
Therefore, the results in Tb:BIG demonstrate the feasibility of the orthogonal MOKE originating from the multipolar structure of Berry curvature in garnet materials.

We now proceed to demonstrate the orthogonal MOKE in ferromagnetic metals \cite{1985_Bader_monoFe, 1983_Buschow_JMMM, 2024_Timo_Cubic}, such as Ni, which are widely recognized as popular magneto-optical materials, together with garnet ferrimagnets. Figure \ref{fig3}a illustrates the measurement setup of the orthogonal MOKE in a 87-nm fcc Ni(111) film.
Figure \ref{fig3}b presents the field sweeping results of Kerr rotation at different directions of \(\boldsymbol{H}\), showing oscillating hysteresis loops similar to those observed in Tb:BIG. 
The asymmetric components $\theta^A_k$ at different $\psi$, shown in Fig. \ref{fig3}c, are also well fitted with the function $a\cdot \cos 3\psi$, confirming the same $\bm M$-multipole Berry curvature origin as in Tb:BIG, despite the difference between ferrimagnetic insulators and ferromagnetic metals. The observations in Ni and Tb:BIG reveal the common existence of $\bm M$-multipole-induced orthogonal MOKE in conventional cubic ferromagnets.


Finally, we demonstrate orthogonal MOKE in a van der Waals ferromagnet, a material family that heavily relies on MOKE for magnetic characterization \cite{2017_Zhang_Cr2Ge2Te6, 2017_Xu_CrI3, 2018_Xu_F3GT, 2019_Xu_F5GT, 2021_Parkin_CrCl3} and exhibits emergent magneto-optical phenomena \cite{2024_Zhang_NP}.
Here we investigate the orthogonal MOKE in \(\rm Fe_5GeTe_2\), a typical van der Waals ferromagnet with lower crystal symmetry than cubic lattice.
Figure \ref{fig4}a illustrates the atomic structure of \(\rm Fe_5GeTe_2\) along $c$ axis. 
Figure \ref{fig4}b shows the 88-nm \(\rm Fe_5GeTe_2\) sample used in experiment, with an in-plane saturation field smaller than 100 Oe at 200 K. 
The sample is rotated in the plane to change the relative orientation $\varphi$ to the magnetic field.
When $\varphi=0^\circ$, no magnetic field dependence in Kerr rotation is observed (Figure \ref{fig4}c). 
However, when $\varphi=90^\circ$, a pronounced hysteresis loop appears, indicating the in-plane magnetization (Figure \ref{fig4}d). 
To assess any longitudinal MOKE contribution from potential imperfect normal reflection, the sample is rotated 180 degrees while maintaining its inclination with a gradienter. 
After rotation, no magnetization dependence is observed when $\varphi=180^\circ$ (Figure \ref{fig4}f), while an inverted hysteresis loop appears when $\varphi=270^\circ$ (Figure \ref{fig4}e), confirming the presence of the orthogonal MOKE signal in \(\rm Fe_5GeTe_2\) sample. Given the symmetry breaking in \(\rm Fe_5GeTe_2\), the $\bm M$-dipole contribution is shown to be incapable of generating the orthogonal MOKE, as elaborated in Section III of the supplementary material. 
Therefore, the detected signal must originate exclusively from the $\bm M$-multipole of Berry curvature. 

We are now in a position to outline the general symmetry requirements for enabling the orthogonal MOKE geometry. 
The symmetric dipole contribution of Berry curvature in $\bm M$ space requires that the eigenvalues are not all equal while the sample plane is not the mirror of the equivalent ellipsoid.
When considering the asymmetric dipole contribution, the orthogonal geometry necessitates that the sample retains at most a mirror reflection symmetry, the plane of which is not parallel to the sample surface, or a $C_n$ rotation symmetry, the axis of which is not perpendicular to the sample surface \cite{2024_Pan_PSHE}, as in the vicinal interface sensitive MOKE \cite{1999_Rasing_Vicinal, 2003_Hamrle_Vicinal}. 
In contrast, the orthogonal geometry in the case of multipole contributions only requires the absence of both $C_{2z}$ and $M_z$ symmetries in the sample, where $z$ denotes the direction perpendicular to the sample surface, being much less restrictive.
For instance, the orthogonal MOKE can typically be expected in single-crystalline samples with a cubic lattice, except for crystals oriented along (001) or (011) planes, which preserve the $C_{2z}$ symmetry \cite{1997_Fert_Quad}.
In polycrystalline samples, however, the orthogonal MOKE signals from individual grains cancel each other out, rendering the overall signal undetectable in transverse MOKE measurements \cite{2008_Allwood_APL}, where the Poynting vector is orthogonal to the magnetization. 
For crystals with lower symmetry than cubic, the conditions for orthogonal MOKE are more easily satisfied, broadening its observability. For example, van der Waals ferromagnets such as $\rm{Cr_2Ge_2Te_6}$ \cite{2017_Zhang_Cr2Ge2Te6}, $\rm{CrI_3}$ \cite{2017_Xu_CrI3}, $\rm{Fe_5GeTe_2}$ \cite{2021_Li_F5GT}, $\rm{Cr_2Te_3}$ \cite{2020_He_Cr2Te3}, $\rm{CrTe_2}$ \cite{2020_Zhang_CrTe2}, and $\rm{CrCl_3}$ \cite{2021_Parkin_CrCl3} all meet these criteria. The in-plane anomalous Hall signal observed in $\rm CrTe_2$ confirms this assertion \cite{2024_PWZ_IPAHE}. Our measurements of the orthogonal MOKE in Bi:TIG, Ni, and $\rm{Fe_5GeTe_2}$ provide comprehensive verification of this symmetry guideline, underscoring orthogonal MOKE as a powerful technique for characterizing magnetization and crystal orientation in van der Waals ferromagnets at the microscale.


The orthogonal MOKE can be readily implemented in various magneto-optical measurements, introducing new functionalities that enhance traditional techniques. For instance, it enables Kerr microscope to monitor the domain structures in ferromagnetic films with in-plane magnetization normally incident light \cite{2012_Liu_IPswitching, 2021_Parkin_CrCl3,2006_GaMnAs}. Additionally, the orthogonal MOKE introduces in-plane magnetization sensing capabilities to the Sagnac interferometer MOKE, which operates solely through normally reflected light and achieves nanoradian sensitivity\cite{2006_Sagnac}. Although the orthogonal MOKE signal may contribute only a fraction of the longitudinal signal, e.g., $29.8\%$ for Kerr rotation in Tb:BIG, the three-order improvement in measurement accuracy provided by the Sagnac MOKE could make the orthogonal MOKE signal a superior probe for ultra-weak in-plane magnetic moments, which are crucial for understanding the spin generation in nonmagnetic materials as in the spin Hall effect and orbital Hall effect \cite{2015_Sinova_SHE,2023_Choi_OHE}.

Beyond the immediate application of the orthogonal MOKE in magnetization characterization, the exploration of additional magneto-optical effects originating from magnetization multipoles holds great promise. Firstly, the orthogonal geometry of magneto-optical effect across different wavelengths—such as microwave, terahertz, and infrared—is ready for investigation, given their shared origin in Berry curvature \cite{1952_microwave_Faraday,2016_THz}. Secondly, the orthogonal geometry of the Faraday effect is highly anticipated due to its common physical foundation with the MOKE\cite{1955_Argyres_Theory}, potentially giving rise to a plethora of new optical devices. For example, if a ferromagnet with in-plane magnetic anisotropy can demonstrate significant Faraday rotation in the orthogonal geometry, it becomes feasible to realize an orthogonal magnetic wave plate or Faraday isolator that does not require an external magnetic field \cite{2023_Faraday_isolator}. The Faraday rotation of such an orthogonal device can be tuned via the direction of the in-plane magnetization, providing an additional degree of freedom not available in conventional Faraday rotators \cite{2023_Faraday_isolator}. Overall, the advent of orthogonal magneto-optical phenomena opens new frontiers in both fundamental physics and technological innovation, paving the way for inventing magneto-optical devices with unprecedented functionality and tunability.


\section*{Methods}

\noindent$\textbf{Sample growth and characterization}$.

The micrometer-thin Tb:BIG film was prepared by the standard liquid phase epitaxy (LPE) method from $Bi_2O_3$- $Li_2CO_3$ based high-temperature solutions at 855 $^{\circ}C$. The Ca, Mg, Zr substituted gadolinium gallium garnet (sGGG, 111, Saint-Gobain) was used as the substrate. Firstly, the high-purity powders (99.999\% metal basis) of $Tb_4O_7$, $Bi_2O_3$, $Fe_2O_3$ and $Li_2CO_3$ were thoroughly mixed in a platinum crucible. In this case, the ratio of powders was 1:86:11:5.
The mixture was melted at 1000 $^{\circ}C$ in the vertical tube LPE furnace, stirred with a platinum rotator until uniform. Then the solution was rapidly cooled at 100 $^{\circ}C$/h down to the growth temperature and maintained a supersaturation state. Next, the substrates horizontally dipped into melt and initiated epitaxy. During the epitaxy, the substrate rotated with rate of 100 rpm for 1 min. After epitaxy, the sample was pulled out from the solution and rotated at 300 rpm to get rid of the liquid solution remnants. Finally, nitric acid cleaning was performed on the sample to obtain a smooth surface.

The Tb:BIG sample is characterized by X-ray diffraction (XRD), vibrating sample magnetometer (VSM) and Faraday effect, as shown in Fig.\ref{fig2}b and extended Fig.\ref{exfig1}, which exhibits high-quality monocrystallinity and in-plane magnetic anisotropy. The longitudinal MOKE signal is measured, as shown in extended Fig. \ref{exfig2}c.

The $87\ \rm nm$ Ni(111) thin film was grown on MgO(111) substrate by DC magnetic sputtering \cite{1998_Ni_growth} at the temperature of $350\ \rm ^\circ C$, with pre-annealing at $600\ \rm ^\circ C$ for $1\ \rm h$ and post-annealing at $300\ \rm ^\circ C$ for $0.5\ \rm h$.
A 4 nm Al coating layer was deposited on the Ni film at room temperature and oxidized in the atmosphere.
The Ar pressure and DC power were set to $5.0\times10^{-3}\rm\ mbar$ and $50\ \rm W$.
The XRD and VSM characterization results are shown in extended Fig.\ref{exfig3}. 

The growth and characterization of bulk $\rm Fe_5GeTe_2$ is described in the literature by Peng Li\cite{2021_Li_F5GT}. The $\rm Fe_5GeTe_2$ sample was exfoliated onto an oxidized silicon substrate. The thickness is determined as 88 nm by atomic force microscopy, as shown in extended Fig. \ref{exfig5}a. The longitudinal MOKE result is shown in extended Fig. \ref{exfig5}b, with experiment setup shown in extended Fig. \ref{exfig5}c. 

\noindent$\textbf{MOKE measurement setup}$. 

In the MOKE measurements in Tb:BIG and Ni, we use the Toptica iBeam Smart laserset as the source of $447\rm\ nm$ light. 
The full optical path is shown in extended Fig. \ref{exfig6}a.
A photoelastic modulator is used to modulate the intensity of light from laserset at $100\rm\ kHz$.
Then, the light is polarized by a Glan prism and Brewster window. To ensure that the incident light is perpendicular to the sample surface, an iris diaphragm (ID1) with a diameter of 0.8 mm is placed in the path of light at a distance of about 145 mm from the sample. When the reflected light is adjusted to pass through ID1, the misaligned angle from normal incidence can be controlled within $\rm 0.32^\circ$. Another diaphragm ID2 is placed after beam splitter to control the normal reflection during the sample rotation. After reflection from the sample, the rotation and ellipticity of light is measured by a quarter-wave plate (QWP), half-wave plate (HWP), Wollaston prism (WP) and balance detector (BD, Thorlabs PDB210A) \cite{1997_Fert_Quad}. 
The output electric signal of BD is demodulated at $100\rm\ kHz$ by two lock-in amplifiers (CIQ Melab, SINE OE1022) to get the absolute value of Kerr rotation and ellipticity. The signs of Kerr rotation ($\theta_k$) and Kerr ellipticity ($\varepsilon_k$) are defined in accordance with the Stokes vector \cite{1997_Zvezdin_Book}. 

The Kerr microscopy used in this study is a commercial MOKE microscopy system from TUOTUO, which has a $450\rm\ nm$ LED as the light source, an in-plane electromagnets and a wet cooling system using liquid nitrogen. The orthogonal geometry measurement is shown in extended Fig. \ref{exfig6}b. The light is polarized and detected by wire grid polarizers. The magnetic field is applied in the sample plane. For the longitudinal measurement in Kerr microscopy, an off-axis diaphragm is inserted to achieve oblique incidence of light, as shown in extended Fig. \ref{exfig5}.

\noindent$\textbf{Details of data analysis}$. 

The asymmetric Kerr rotation ($\theta^A_k$) in Tb:BIG and Ni sample is defined as $(\theta^+_k-\theta^-_k)/2$, where $\theta^+_k$ and $\theta^-_k$ are the intercepts of Kerr signal in the positive and negative saturated regions, obtained through linear fitting, as shown in Fig. \ref{fig2}c. 
Similarly, the asymmetric Kerr ellipticity $\varepsilon^A_k$ is defined as $(\varepsilon^+_k-\varepsilon^-_k)/2$. 
The saturated regions used for linear fitting is $|H|>100\ \rm Oe$ for the Tb:BIG sample and $|H|>120\ \rm Oe$ for the Ni sample.

The Orthogonal MOKE signal of Tb:BIG around zero field is explained by the deviation of $\bm M$ from $\bm H$ direction with additional quadratic MOKE contribution. 
In the measurement of Ni, the non-zero residual magnetization and the change in polarization with respect to $\bm H$ contribute a butterfly-shape quadratic MOKE signal around zero field.

\bibliographystyle{Science}
\bibliography{CMOKE}

\begin{thebibliography}{10}

\bibitem{2000_Qiu_MOKEreview}
Z.~Q. Qiu, S.~D. Bader, {\it Rev. Sci. Instrum.\/} {\bf 71}, 1243 (2000).

\bibitem{1984_MODisc}
K.~Schouhamer~Immink, J.~Braat, {\it J. Audio Eng. Soc.\/} {\bf 32} (1984).

\bibitem{2003_MOrecording}
D.~Jenkins, {\it et~al.\/}, {\it Microsyst. Technol.\/} {\bf 10}, 66 (2003).

\bibitem{1877_Kerr}
J.~Kerr, {\it The London, Edinburgh, and Dublin Philosophical Magazine and
  Journal of Science\/} {\bf 3}, 321 (1877).

\bibitem{1878_Kerr}
J.~Kerr, {\it The London, Edinburgh, and Dublin Philosophical Magazine and
  Journal of Science\/} {\bf 5}, 161 (1878).

\bibitem{1954_Edward_LMOKE}
C.~A. Fowler, E.~M. Fryer, {\it Phys. Rev.\/} {\bf 94}, 52 (1954).

\bibitem{1968_Freiser_MOreview}
M.~Freiser, {\it IEEE Trans. Magn.\/} {\bf 4}, 152 (1968).

\bibitem{1990_Zak_MO}
J.~Zak, E.~Moog, C.~Liu, S.~Bader, {\it J. Magn. Magn. Mater.\/} {\bf 89}, 107
  (1990).

\bibitem{1997_Zvezdin_Book}
A.~K. Zvezdin, V.~A. Kotov, {\it Modern magnetooptics and magnetooptical
  materials\/} (CRC Press, 1997).

\bibitem{2006_Nagaosa_AHE}
N.~Nagaosa, {\it J. Phys. Soc. Jpn.\/} {\bf 75}, 042001 (2006).

\bibitem{2010_MacDonald_AHE}
N.~Nagaosa, J.~Sinova, S.~Onoda, A.~H. MacDonald, N.~P. Ong, {\it Rev. Mod.
  Phys.\/} {\bf 82}, 1539 (2010).

\bibitem{2015_Hou_AHE}
D.~Hou, {\it et~al.\/}, {\it Phys. Rev. Lett.\/} {\bf 114}, 217203 (2015).

\bibitem{1999_Rasing_Vicinal}
A.~V. Petukhov, A.~Kirilyuk, T.~Rasing, {\it Phys. Rev. B\/} {\bf 59}, 4211
  (1999).

\bibitem{2003_Hamrle_Vicinal}
J.~Hamrle, {\it et~al.\/}, {\it Phys. Rev. B\/} {\bf 67}, 155411 (2003).

\bibitem{2024_WU_IPAHE}
L.~Wang, {\it et~al.\/}, {\it Phys. Rev. Lett.\/} {\bf 132}, 106601 (2024).

\bibitem{2024_PWZ_IPAHE}
W.~Peng, {\it et~al.\/}, {\it arXiv:2402.15741\/}  (2024).

\bibitem{2024_ZhengLiu_Multipolar}
Z.~Liu, M.~Wei, D.~Hou, Y.~Gao, Q.~Niu, {\it arXiv:2408.08810\/}  (2024).

\bibitem{2019_nonlinearHall}
Q.~Ma, {\it et~al.\/}, {\it Nature\/} {\bf 565}, 337 (2019).

\bibitem{2015_Fu_PRL}
I.~Sodemann, L.~Fu, {\it Phys. Rev. Lett.\/} {\bf 115}, 216806 (2015).

\bibitem{2021_Qiong_NM}
Q.~Ma, A.~G. Grushin, K.~S. Burch, {\it Nat. Mater.\/} {\bf 20}, 1601 (2021).

\bibitem{2004_Yao}
Y.~Yao, {\it et~al.\/}, {\it Phys. Rev. Lett.\/} {\bf 92}, 037204 (2004).

\bibitem{2024_Xu_viewpoint}
S.-Y. Xu, {\it Physics\/} {\bf 17}, 38 (2024).

\bibitem{2021_GarnetApplication}
Y.~Yang, T.~Liu, L.~Bi, L.~Deng, {\it J. Alloys Compd.\/} {\bf 860}, 158235
  (2021).

\bibitem{2023_Faraday_isolator}
L.~Zhang, {\it et~al.\/}, {\it J. Adv. Ceram.\/} {\bf 12}, 873 (2023).

\bibitem{1985_Bader_monoFe}
E.~Moog, S.~Bader, {\it Superlattices Microstruct.\/} {\bf 1}, 543 (1985).

\bibitem{1983_Buschow_JMMM}
K.~Buschow, P.~{van Engen}, R.~Jongebreur, {\it J. Magn. Magn. Mater.\/} {\bf
  38}, 1 (1983).

\bibitem{2024_Timo_Cubic}
M.~Gaerner, R.~Silber, T.~Peters, J.~Hamrle, T.~Kuschel, {\it Phys. Rev.
  Appl.\/} {\bf 22}, 024066 (2024).

\bibitem{2017_Zhang_Cr2Ge2Te6}
C.~Gong, {\it et~al.\/}, {\it Nature\/} {\bf 546}, 265 (2017).

\bibitem{2017_Xu_CrI3}
B.~Huang, {\it et~al.\/}, {\it Nature\/} {\bf 546}, 270 (2017).

\bibitem{2018_Xu_F3GT}
Z.~Fei, {\it et~al.\/}, {\it Nat. Mater.\/} {\bf 17}, 778 (2018).

\bibitem{2019_Xu_F5GT}
A.~F. May, {\it et~al.\/}, {\it ACS Nano\/} {\bf 13}, 4436 (2019).

\bibitem{2021_Parkin_CrCl3}
A.~Bedoya-Pinto, {\it et~al.\/}, {\it Science\/} {\bf 374}, 616 (2021).

\bibitem{2024_Zhang_NP}
X.~Li, {\it et~al.\/}, {\it Nat. Phys.\/} {\bf 20}, 1145 (2024).

\bibitem{2024_Pan_PSHE}
H.~Pan, Z.~Liu, D.~Hou, Y.~Gao, Q.~Niu, {\it Phys. Rev. Res.\/} {\bf 6},
  L012034 (2024).

\bibitem{1997_Fert_Quad}
K.~Postava, {\it et~al.\/}, {\it J. Magn. Magn. Mater.\/} {\bf 172}, 199
  (1997).

\bibitem{2008_Allwood_APL}
D.~A. Allwood, {\it et~al.\/}, {\it Appl. Phys. Lett.\/} {\bf 92}, 072503
  (2008).

\bibitem{2021_Li_F5GT}
L.~Alahmed, {\it et~al.\/}, {\it 2D Mater.\/} {\bf 8}, 045030 (2021).

\bibitem{2020_He_Cr2Te3}
Y.~Wen, {\it et~al.\/}, {\it Nano Lett.\/} {\bf 20}, 3130 (2020).

\bibitem{2020_Zhang_CrTe2}
X.~Sun, {\it et~al.\/}, {\it Nano Res.\/} {\bf 13}, 3358 (2020).

\bibitem{2012_Liu_IPswitching}
L.~Liu, {\it et~al.\/}, {\it Science\/} {\bf 336}, 555 (2012).

\bibitem{2006_GaMnAs}
M.~Yamanouchi, D.~Chiba, F.~Matsukura, T.~Dietl, H.~Ohno, {\it Phys. Rev.
  Lett.\/} {\bf 96}, 096601 (2006).

\bibitem{2006_Sagnac}
J.~Xia, Y.~Maeno, P.~T. Beyersdorf, M.~M. Fejer, A.~Kapitulnik, {\it Phys. Rev.
  Lett.\/} {\bf 97}, 167002 (2006).

\bibitem{2015_Sinova_SHE}
J.~Sinova, S.~O. Valenzuela, J.~Wunderlich, C.~H. Back, T.~Jungwirth, {\it Rev.
  Mod. Phys.\/} {\bf 87}, 1213 (2015).

\bibitem{2023_Choi_OHE}
Y.-G. Choi, {\it et~al.\/}, {\it Nature\/} {\bf 619}, 52 (2023).

\bibitem{1952_microwave_Faraday}
C.~L. Hogan, {\it The Bell System Technical Journal\/} {\bf 31}, 1 (1952).

\bibitem{2016_THz}
J.~Walowski, M.~Münzenberg, {\it J. Appl. Phys.\/} {\bf 120}, 140901 (2016).

\bibitem{1955_Argyres_Theory}
P.~N. Argyres, {\it Phys. Rev.\/} {\bf 97}, 334 (1955).

\bibitem{1998_Ni_growth}
P.~Sandström, E.~B. Svedberg, J.~Birch, J.-E. Sundgren, {\it MRS Online
  Proceedings Library\/} {\bf 528}, 209 (1998).

\end{thebibliography}
\newpage
\noindent$\textbf{Acknowledgments}$: The author acknowledges Timo Kuschel for valuable discussion. This work was supported by the National Key R\&D Program under grant Nos. 2022YFA1403502, 2023YFA1406400, the National Natural Science Foundation of China (12234017, 12074366, 12174364), the Fundamental Research Funds for the Central Universities (Grant No. WK9990000116). Z. Liu is supported by the National Natural Science Foundation of China (11974327 and 12004369), Fundamental Research Funds for the Central Universities (WK3510000010, WK2030020032), Anhui Initiative in Quantum Information Technologies (No. AHY170000), and Innovation Program for Quantum Science and Technology (2021ZD0302800). The setup of magnetic sputtering system and thin film growth was assisted by Anhui epitaxy technology co. Ltd. The sample fabrication was supported by the USTC Center for Micro- and Nanoscale Research and Fabrication. 

\noindent$\textbf{Author contributions}$: H. Pan and D. Hou designed the experiment. H. Li and Q. Yang grew the Tb:BIG sample. H. Pan grew the Ni sample. J. Huang and P. Li fabricated the $\rm Fe_5GeTe_2$ sample. H. Pan , M. Fang and W. Peng performed characterization of the Ni and Tb:BIG sample. H. Pan and M. Fang carried out the MOKE measurements in the Ni and Tb:BIG sample. X. Hu, X. Chang and Z. Sheng tested the magneto-optical property of the $\rm Fe_5GeTe_2$ sample. Y. Jia and L. Wang provided Bi:YIG sample for magneto-optical testing. H. Pan, Y. Yuan, D. Liu and Q. Li carried out the MOKE measurement in the $\rm Fe_5GeTe_2$ sample. H. Pan, X. Chen and D. Hou analyzed the experiment data. Z. Liu, D. Hou, Q. Niu and Y. Gao developed the explanation of the experiment. H. Pan and D. Hou wrote the manuscript. All authors discussed the results and commented on the manuscript.

\noindent$\textbf{Competing interests}$: The authors declare that they have no competing interests.

\linespread{1.5}

\newcounter{Figure}
\setcounter{Figure}{1}
\renewcommand{\thefigure}{\arabic{Figure}}
\renewcommand{\figurename}{Figure}
\newpage
\begin{figure}[ht]
        \centering
        \makebox[\textwidth][c]{\includegraphics[width=172mm]{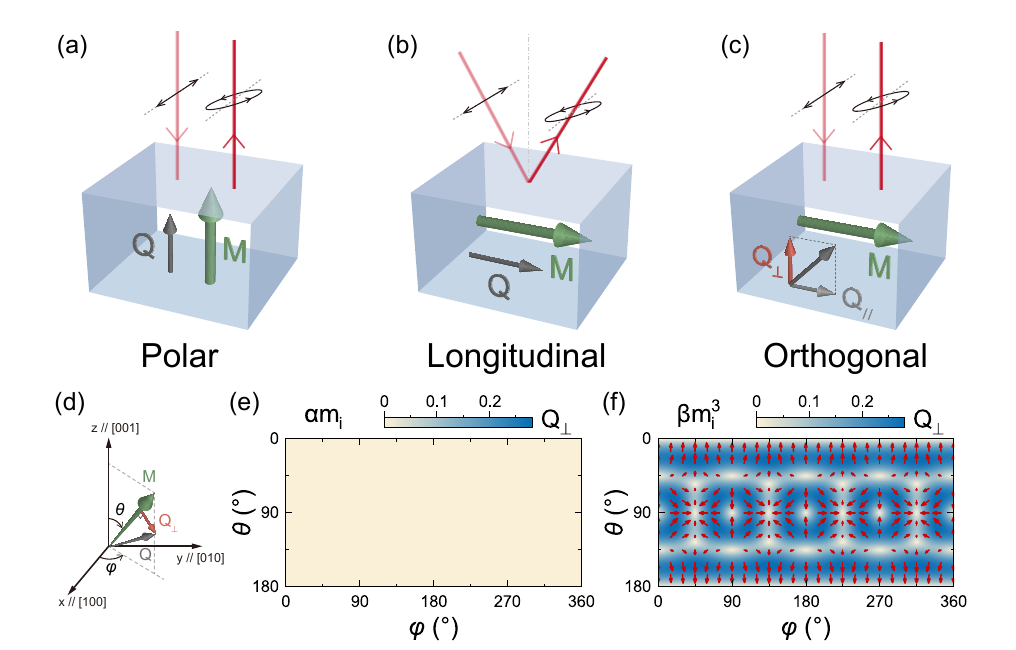}}
        \caption{\textbf{Geometries of magneto-optical Kerr effect (MOKE).}
\textbf{a-c}, Schematic diagrams illustrating the three different MOKE geometries. Green and gray arrows represent the magnetization (\(\boldsymbol{M}\)) of material and the induced Voigt vector (\(\boldsymbol{Q}\)). The perpendicular component of $\bm Q$ relative to $\bm M$, denoted as \(\boldsymbol{Q}_\bot\), is highlighted by a red arrow. The incident lights (light red lines) are linearly polarized in the direction indicated by black arrows, while the reflected lights (dark red lines) from the sample surface (\(\boldsymbol{S}\)) carry Kerr signals, which consist of Kerr rotation and Kerr ellipticity, depicted by black ellipses. 
\textbf{a}, Polar MOKE geometry, where \(\boldsymbol{M},\boldsymbol{Q}\ \bot\ \boldsymbol{S}\), inducing Kerr signal in normally reflected light.
\textbf{b}, Longitudinal MOKE geometry, where \(\boldsymbol{M},\boldsymbol{Q}\ \parallel \ \boldsymbol{S}\), inducing Kerr signal in obliquely reflected light.
\textbf{c}, Orthogonal MOKE, with the \(\boldsymbol{Q}_\bot\) perpendicular to the in-plane \(\boldsymbol{M}\), resulting in Kerr signal in normal reflection.
\textbf{d}, \(\boldsymbol{M}\), \(\boldsymbol{Q}\) and \(\boldsymbol{Q}_\bot\) components in cubic lattice. The \(x,y,z\) axes are parallel to [100],[010],[001],respectively. The direction of \(\boldsymbol{M}\) is defined by spherical coordinates, \(\theta\) and \(\varphi\). 
\textbf{e-f}, The magnitude and direction of \(\boldsymbol{Q_\bot}\) arising from $\alpha m_i$ (\textbf{e}) and $\beta m^3_i$ (\textbf{f}) terms in Eq. \ref{equ:cubic}. 
}
        \label{fig1}  
\end{figure}

\newpage
\setcounter{Figure}{2}
\begin{figure}[ht]
        \centering
        \makebox[\textwidth][c]{\includegraphics[width=180mm]{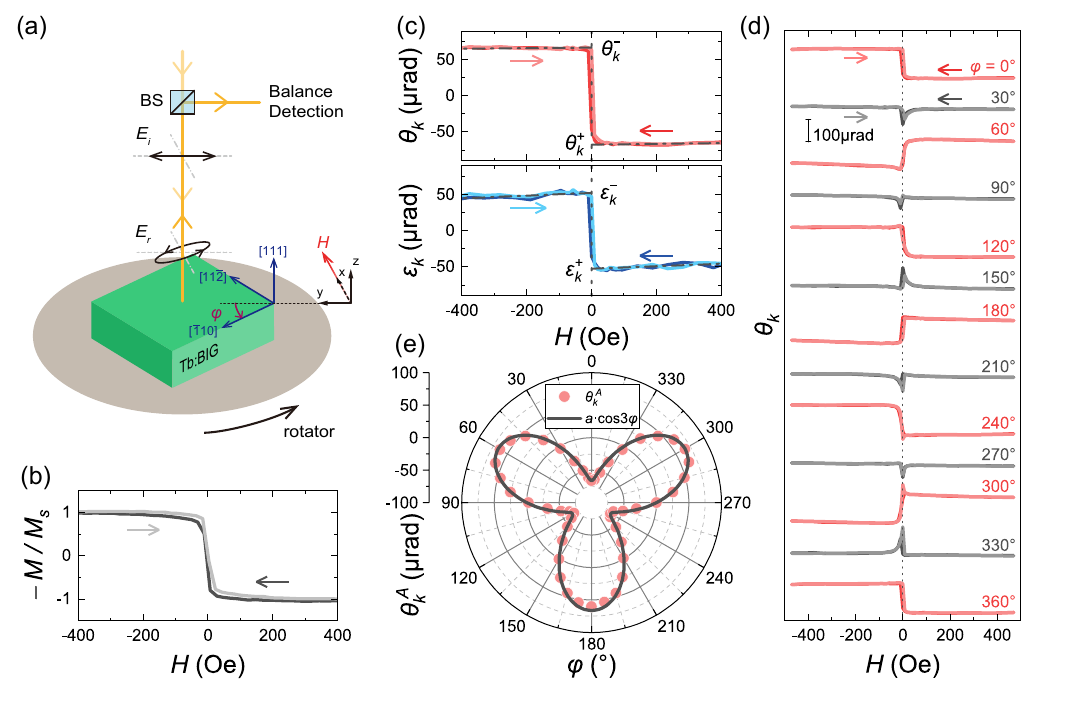}}
        \caption{\textbf{MOKE signal of a $\bf Tb_{2.22}Bi_{0.78}Fe_{5}O_{12}$ (Tb:BIG) in orthogonal geometry.} 
\textbf{a}, Schematic diagram of the orthogonal MOKE measurement setup for a Tb:BIG(111) film, which is rotated in $xy$ plane by an angle $\varphi$. The magnetic field (\(\boldsymbol{H}\)) is applied along $x$ direction. A 447-nm laser beam with linear polarization ($E_i$) along $y$ axis is normally incident to the surface of the Tb:BIG sample. The Kerr rotation ($\theta_k$) and Kerr ellipticity ($\varepsilon_k$) of the reflected light are measured by balance detection. 
\textbf{b}, In-plane magnetization hysteresis of the Tb:BIG sample measured by vibrating sample magnetometer (VSM).
\textbf{c}, Field dependence of $\theta_k$ and $\varepsilon_k$ of the Tb:BIG sample in orthogonal geometry when $\varphi=0.0^\circ$. $\theta^+_k$, $\theta^-_k$, $\varepsilon^+_k$ and $\varepsilon^-_k$ denote the intercepts of linear fittings of $\theta_k$ and $\varepsilon_k$ in positive and negative saturated regions. The arrows represent the sweeping directions of magnetic field.
\textbf{d}, Field dependence of $\theta_k$ of the Tb:BIG sample at different sample orientations $\varphi$. 
\textbf{e}, Dependence of $\theta^A_k=(\theta^+_k-\theta^-_k)/2$ on $\varphi$ for the Tb:BIG sample, which is well fitted with a $\cos 3\varphi$ function (gray line).
}
        \label{fig2}  
\end{figure}

\newpage
\setcounter{Figure}{3}
\begin{figure}[ht]
        \centering
        \makebox[\textwidth][c]{\includegraphics[width=80mm]{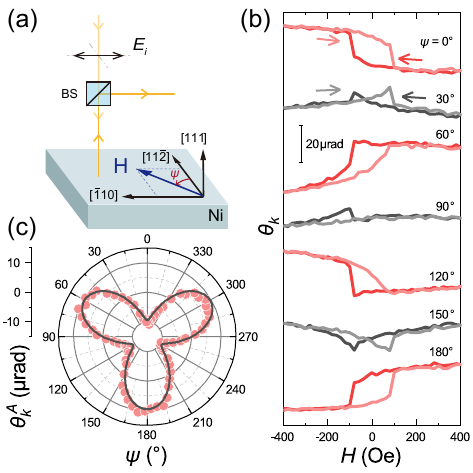}}
        \caption{\textbf{MOKE signal of nickel in orthogonal geometry.} 
\textbf{a,}, Schematic diagram of the orthogonal MOKE measurement in a 87nm-thick Ni(111) film on MgO(111) substrate. The magnetic field is applied in the sample plane at an angle $\psi$ relative to the $[11\bar{2}]$ direction of Ni. A 447-nm laser beam with linear polarization ($E_i$) along $[\bar{1}10]$ is normally incident to the sample surface. The Kerr rotation ($\theta_k$) of the reflected light is measured by balance detection.
{\bf (b)} Field dependence of Kerr rotation of the Ni$(111)$ sample at different $\psi$. The arrows represent the field sweeping directions.
{\bf (c)} $\psi$ dependence of $\theta^A_k$, the asymmetric component of Kerr rotation, which is fitted with a $\cos 3\psi$ function (gray line).
}
        \label{fig3}  
\end{figure}

\newpage
\setcounter{Figure}{4}
\begin{figure}[ht]
        \centering
        \makebox[\textwidth][c]{\includegraphics[width=180mm]{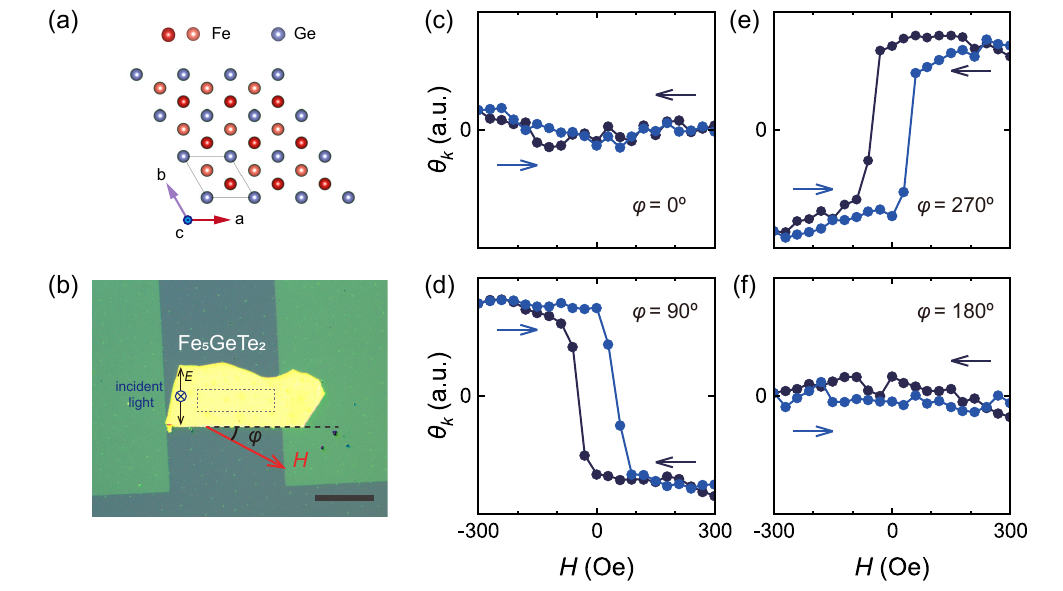}}
        \caption{\textbf{MOKE signal of $\bf Fe_5GeTe_2$ in orthogonal geometry.} 
\textbf{a,} Atomic structure of $\rm Fe_5GeTe_2$ projected along $c$ axis, showing a 3-fold symmetry. 
\textbf{b,} Optical image of the $\rm Fe_5GeTe_2$ sample. The Kerr rotation of the marked area on the sample is measured by Kerr microscopy with normally incident light at 200 K. The angle between the long edge of the sample and the in-plane magnetic field is defined as $\varphi$. The black arrow denotes the polarization of light. The length of the scale bar is 50 $\rm\mu m$.
\textbf{c -- f,} \(\boldsymbol{H}\) dependence of Kerr rotation with four different orientations of the sample with respect to the in-plane \(\boldsymbol{H}\). 
}
        \label{fig4}  
\end{figure}

\newcounter{Exfigure}
\setcounter{Exfigure}{1}
\renewcommand{\thefigure}{\arabic{Exfigure}}
\renewcommand{\figurename}{Extended Figure}
\newpage
\begin{figure}[ht]
        \centering
        \makebox[\textwidth][c]{\includegraphics[width=150mm]{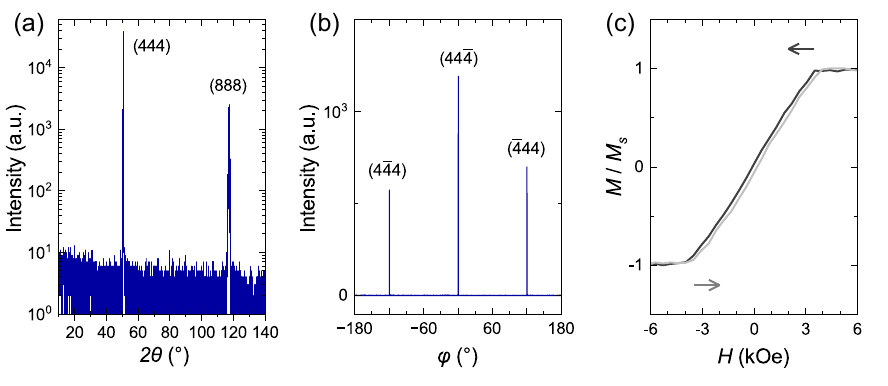}}
        \caption{\textbf{Characterization results of the Tb:BIG sample.}
\textbf{a,} $2\theta-\omega$ scan of the Tb:BIG sample in out-of-plane direction, measured by XRD. 
\textbf{b,} $\varphi$ scan of the Tb:BIG sample with $\chi=71^\circ$ and $2\theta=50.470^\circ$, measured by XRD. 
\textbf{c,} Out-of-plane magnetization hysteresis of the Tb:BIG sample, measured by Faraday effect.
}
        \label{exfig1}  
\end{figure}

\setcounter{Exfigure}{2}
\newpage
\begin{figure}[ht]
        \centering
        \makebox[\textwidth][c]{\includegraphics[width=105mm]{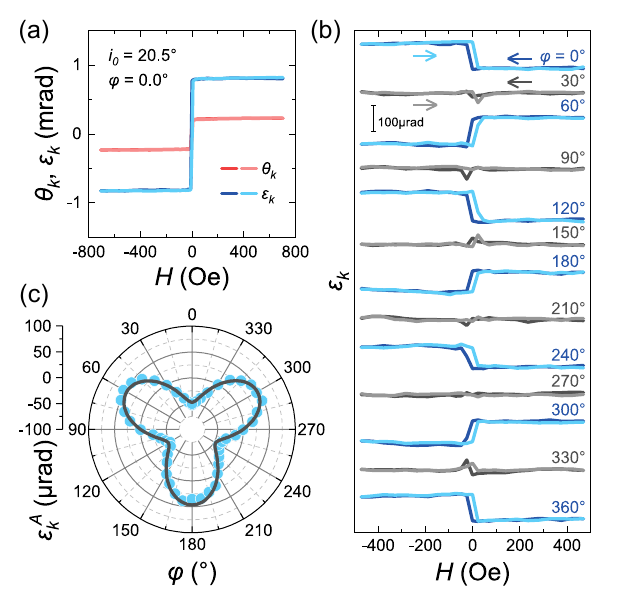}}
        \caption{\textbf{Additional MOKE results of the Tb:BIG sample.}
\textbf{a,} Longitudinal MOKE of the Tb:BIG sample with an incident angle of $20.1^\circ$.
\textbf{b,} Field dependence of $\varepsilon_k$ of the Tb:BIG sample in orthogonal geometry at different sample orientations $\varphi$.
\textbf{c,} $\varphi$ dependence of $\varepsilon^A_k$, the asymmetric Kerr ellipticity of the Tb:BIG sample in orthogonal geometry, which is fitted with a $\cos 3\varphi$ function (gray line).
}
        \label{exfig2}  
\end{figure}

\setcounter{Exfigure}{3}
\newpage
\begin{figure}[ht]
        \centering
        \makebox[\textwidth][c]{\includegraphics[width=133mm]{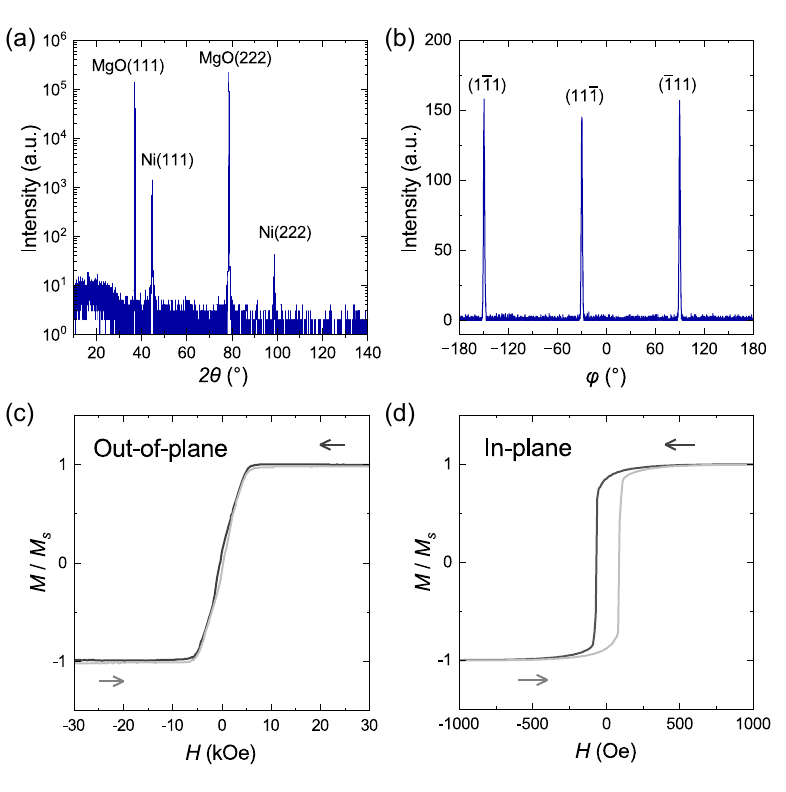}}
        \caption{\textbf{Characterization results of the Ni sample.}
\textbf{a,} $2\theta-\omega$ scan of the Ni sample in out-of-plane direction, measured by XRD. 
\textbf{b,} $\varphi$ scan of the Ni sample with $\chi=71^\circ$ and $2\theta=44.576^\circ$, measured by XRD. 
\textbf{c,} Out-of-plane magnetization hysteresis of the Ni sample, measured by VSM.
\textbf{d,} In-plane magnetization hysteresis of the Ni sample, measured by VSM.
}
        \label{exfig3}  
\end{figure}

\setcounter{Exfigure}{4}
\newpage
\begin{figure}[ht]
        \centering
        \makebox[\textwidth][c]{\includegraphics[width=130mm]{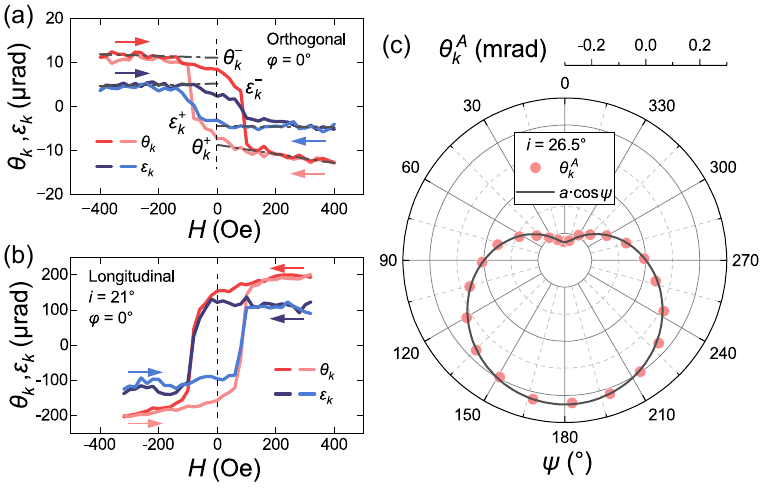}}
        \caption{\textbf{Additional MOKE results of the Ni sample.}
\textbf{a,} Orthogonal MOKE of the Ni sample at $\varphi=0^\circ$.
\textbf{b,} Longitudinal MOKE of the Ni sample with an incident angle of $21^\circ$.
\textbf{c,} $\psi$ dependence of asymmetric Kerr rotation, $\theta^A_k$, in longitudinal geometry with an incident angle of $26.5^\circ$, which is fitted with a $\cos \psi$ function (gray line).
}
        \label{exfig4}  
\end{figure}

\setcounter{Exfigure}{5}
\newpage
\begin{figure}[ht]
        \centering
        \makebox[\textwidth][c]{\includegraphics[width=130mm]{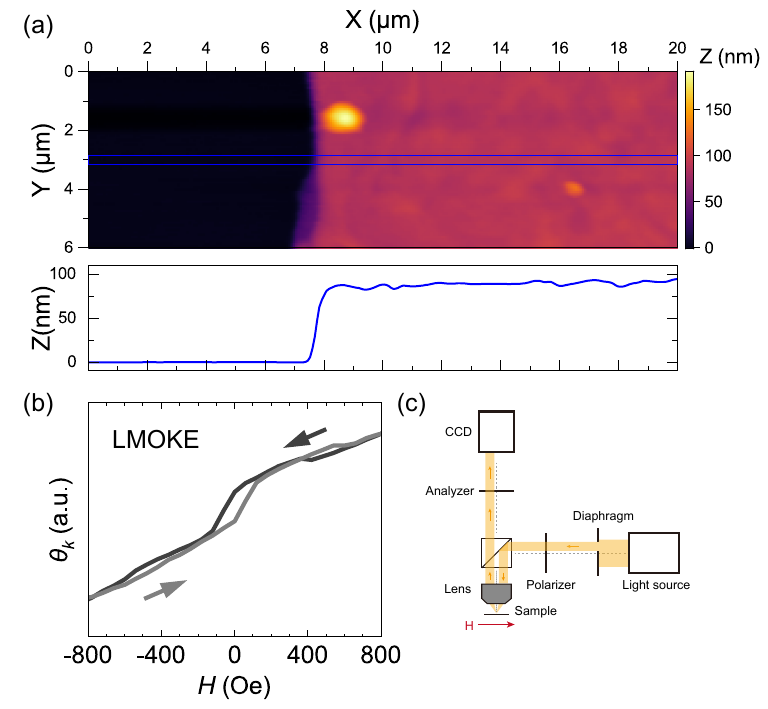}}
        \caption{\textbf{Atomic force Microscopy and longitudinal MOKE results of the $\bf Fe_5GeTe_2$ sample.}
\textbf{a,} AFM image at the edge of the $\rm Fe_5GeTe_2$ sample, including a line profile. The thickness is approximately 88 nm.
\textbf{b,} Longitudinal MOKE of the $\rm Fe_5GeTe_2$ sample measured by Kerr microscopy.
\textbf{c,} Schematic diagram of Kerr microscopy measurement in longitudinal geometry. An off-axis diaphragm is added to achieve longitudinal measurement.
}
        \label{exfig5}  
\end{figure}

\setcounter{Exfigure}{6}
\newpage
\begin{figure}[ht]
        \centering
        \makebox[\textwidth][c]{\includegraphics[width=120mm]{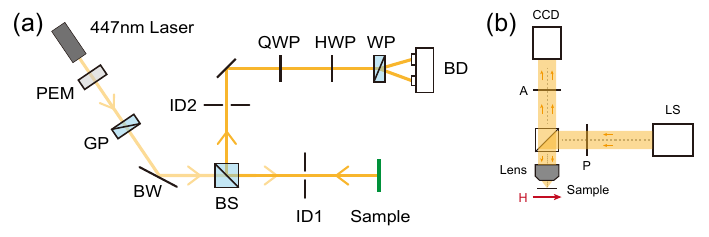}}
        \caption{\textbf{Schematic diagram of optical path of MOKE measurement.} 
\textbf{a,} Laser MOKE measurement setup. The 447-nm laser beam is modulated in amplitude by a photoelastic modulator (PEM) and polarized by a Glan prism (GP) and Brewster window (BW). The polarization of the reflected light is detected by quarter-wave plate (QWP), half-wave plate (HWP) and detected by a balance detector (BD) following a Wollaston prism (WP). The normal reflection is controlled by iris diaphragms (ID1 and ID2).
\textbf{b,} Kerr microscopy measurement setup in Orthogonal geometry. The 450-nm light from light source (LS) is polarized by polarizer (P). The polarization of reflected light is detected by an a (A) and a CCD camera. 
}
        \label{exfig6}  
\end{figure}

\end{document}